\documentclass[aps,twocolumn,preprintnumbers,groupedaddress,nofootinbib,prl]{revtex4-2}
\pdfoutput=1

\newcommand{\SMfilename}{chaotic-kummer-SM}
\def\arXivversion{}

\usepackage{lineno}
\usepackage{amssymb,amsmath,amsthm}
\usepackage[shortlabels]{enumitem}
\usepackage{braket}
\usepackage{comment}
\usepackage{subcaption}
\DeclareCaptionJustification{block}{\leftskip=0pt\rightskip=0pt\parfillskip=0pt plus 1fil\relax}
\usepackage{graphicx, xcolor, varwidth}
\usepackage{color}
\usepackage{csquotes} 
\usepackage{mathtools}
\usepackage{tikz}
\usepackage{commath}
\usepackage{adjustbox}
\usepackage{booktabs}
\usepackage{placeins}
\usepackage{xr-hyper}
\usepackage{hyperref}
\usepackage{dsfont}

\hypersetup{
    pdfstartview={FitH},
    pdftitle={Onset of quantum chaos in a family of two-dimensional conformal field theories},
    pdfauthor={Felix M. Haehl, Moshe Rozali},
    colorlinks=true,
    linkcolor=cyan,
    citecolor=cyan,
    filecolor=magenta,
    urlcolor=blue
}

\makeatletter
\renewcommand\paragraph{\@startsection{paragraph}{4}{\parindent}{\z@}{-1em}{\normalfont\normalsize\bfseries}}
\makeatother

\newcommand{\Sym}{\mathrm{Sym}}
\newcommand{\Tf}{\mathbb T^4}
\newcommand{\Def}{\mathcal D} 
\newcommand{\B}{\mathcal B}
\newcommand{\HB}{H_{\B}}

\newcommand{\cMat}{{\rm c}}
\newcommand{\twist}{{\boldsymbol\sigma}}
\newcommand{\lam}{\lambda}
\newcommand{\Lam}{\Lambda}
\newcommand{\Gt}{\widetilde G}
\newcommand{\rmean}{\langle r\rangle}
\newcommand{\bs}{\mathbf s}
\newcommand{\bt}{\mathbf t}
\newcommand{\bff}{\mathbf f}
\newcommand{\bn}{\mathbf n}
\newcommand{\ba}{\mathbf a}
\newcommand{\bb}{\mathbf b}
\newcommand{\bw}{\mathbf w}

\newcommand{\Dc}{\Def_{\cMat}}
\newcommand{\Mc}{M_{\cMat}} 
\newcommand{\Hc}{{\cal G}_{\cMat}}
\newcommand{\Dcz}{\Def_{\cMat,0}}

\DeclareMathOperator{\Tr}{Tr}

\DeclareMathOperator{\rank}{rank}

\ifdefined\arXivversion
  \usepackage{docmute}
  \makeatletter
  \let\orig@newlabel\newlabel
  \def\newlabel#1#2{\orig@newlabel{#1}{#2}\orig@newlabel{M-#1}{#2}\orig@newlabel{SM-#1}{#2}}
  \AtBeginDocument{\let\SM@hangfrom\@hangfrom@section \let\SM@cntformat\@sectioncntformat}
  \def\SMsetup{\let\@hangfrom@section\SM@hangfrom \let\@sectioncntformat\SM@cntformat
    \@removefromreset{equation}{section}}
  \makeatother
\else
\fi

\begin{document}

\title{Onset of quantum chaos in a family of two-dimensional conformal field theories}

\author{Felix M.\ Haehl}
\affiliation{School of Mathematical Sciences and STAG Research Centre, University of Southampton, SO17 1BJ, U.K.}
\author{Moshe Rozali}
\affiliation{Department of Physics and Astronomy, University of British Columbia, Vancouver, V6T 1Z1, Canada}

\begin{abstract}
We exhibit random matrix statistics in the scaling dimensions of non-BPS primary operators in a family of two-dimensional conformal field theories. Specifically, we consider the sigma model with K3 target space and $\mathcal N=(4,4)$ supersymmetry in a neighborhood of the Kummer orbifold point $\Tf/\mathbb Z_2$. At the free-fermion radius, the spin-zero spectrum of the $\mathbb Z_2^4$-invariant sector is exactly degenerate at integer energies. The exactly marginal blow-up deformations that preserve this symmetry form a four-dimensional space, whose directions we parametrize by unit quaternions. Along each deformation direction, we compute the first-order splitting of the degenerate levels, i.e., the leading anomalous dimensions as the coupling goes to zero. In sectors where all exact symmetries are resolved, including a hidden $\mathbb Z_2$ at the even levels, the non-zero first-order shifts along a generic direction follow the Gaussian orthogonal ensemble (GOE), as we verify by computing several spectral diagnostics. We compare quantitatively with the well-known singlet deformation, which has enhanced symmetry: even after all its known symmetries are resolved, its first-order spectrum has exact degeneracies, its gap ratio approaches the GOE value only slowly as the energy increases, and its unfolded spectrum lacks the long-range rigidity of the GOE.
\end{abstract}

\maketitle

\section{Introduction}
\label{sec:intro}

Quantum chaos has become a cornerstone of holography, in particular for the emergence of semiclassical gravity and the role of wormhole geometries in the gravitational path integral \cite{Shenker:2013pqa,Maldacena:2015waa,Cotler:2016fpe}. While two-dimensional gravity \cite{Almheiri:2014cka,Maldacena:2016upp} enjoys a dual description in terms of explicit chaotic quantum systems \cite{Sachdev:2015efa,Maldacena:2016hyu,Saad:2019lba}, extending this paradigm to higher dimensions remains challenging. In two dimensions, despite compelling evidence for random matrix universality in holographic CFTs \cite{Cotler:2020ugk,Haehl:2023tkr,DiUbaldo:2023qli,Boruch:2025ilr}, direct spectral tests of random matrix universality in explicit CFTs remain scarce.

In this Letter we describe a continuous family of explicit 2d CFTs in which the first-order anomalous dimensions of non-BPS primaries exhibit random matrix statistics. To our knowledge, these are the first two-dimensional CFTs with this property. We do not assume any ensemble average: the leading anomalous dimensions of the primaries of a single theory repel each other, consistent with random matrix theory (RMT) predictions. 

The well-studied solvable 2d CFTs are rational or integrable, and their spectra are degenerate, arithmetic or Poissonian. One route away from this class is to deform a solvable theory: coupled minimal models admit perturbative IR fixed points with strong evidence for irrationality \cite{Dotsenko:1998gyp,Antunes:2022vtb,Antunes:2024mfb,Antunes:2025erb}, coupled anyon chains \cite{Feiguin:2006ydp} with non-invertible symmetries are argued to do the same \cite{Antunes:2025huk}, and the twist deformation of the D1-D5 symmetric orbifold is expected to take it toward the regime of black-hole microstates \cite{David:2002wn,Avery:2010er}. Whether the deformed spectrum is chaotic in the sense of RMT is a different question. In the case of the K3 sigma model at the orbifold points, the undeformed theory is free, but it admits exactly marginal deformations, which lift the degenerate spectrum in first-order perturbation theory and produce large matrices with non-trivial statistics.

\begin{figure*}[t]
\includegraphics[width=.75\textwidth]{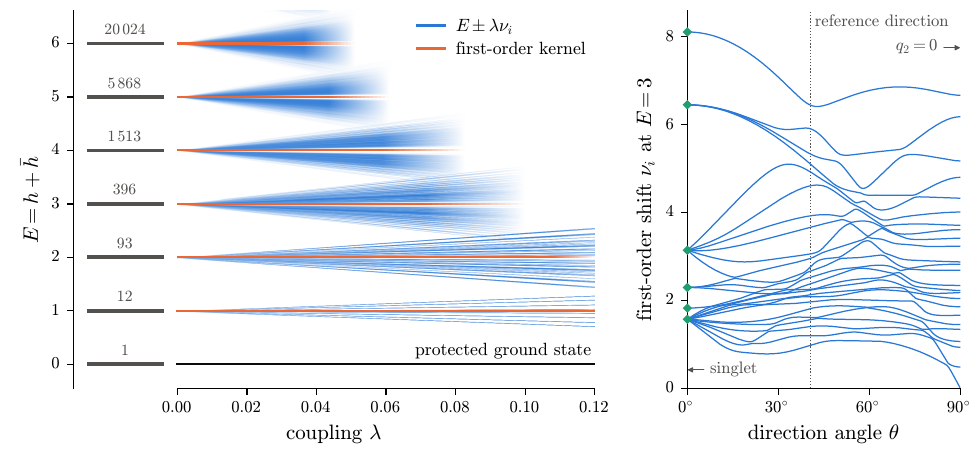}
\caption{The scalar spectrum near the Kummer point.
{\it Left:} At $\lam=0$ the massive primaries of the invariant sector form exactly degenerate blocks at integer $E$ above the protected Ramond ground state (numbers: degeneracies). At first order each block splits into a band $E\pm\lam\nu_i$, while the states with vanishing first-order shift remain degenerate at $E$ (orange).
{\it Right:} Direction dependence in the space of deformations (exact at first order). The $25$ non-zero first-order shifts $\nu_i$ of one fully resolved $E=3$ sub-block, plotted along the great circle $q(\theta)\in S^3$ from the singlet $q_{\rm s}=(0,0,-1,0)$ (green diamonds) through the reference direction (dotted) to the triplet ``mirror'' $q_2=0$.}
\label{fig:spectrum}
\end{figure*}

The K3 sigma model is an $\mathcal N=(4,4)$ superconformal field theory with central charge $c=6$. Its moduli space is the $80$-dimensional Grassmannian $O(4,20)/O(4)\times O(20)$ up to discrete identifications \cite{Seiberg:1999xz,Aspinwall:1996mn,Nahm:1999ps}. At special points in moduli space the model is relatively well understood: at its orbifold points the theory is a free orbifold of a torus model, and at its Gepner points it is rational \cite{Gepner:1987vz,Gepner:1987qi}. The elliptic genus and the $\frac12$-BPS ground states are protected \cite{Dijkgraaf:1996xw,Eguchi:2010ej,Eguchi:1987wf}, while additional $\frac14$-BPS states at special points can lift \cite{Keller:2019suk}. In the D1-D5 system, the $\frac14$-BPS states that survive the deformation have recently been characterized \cite{Benjamin:2021zkn,Chang:2025rqy,Bintanja:2026uwk}, and chaos has even been investigated within the BPS sector itself, where the states are exactly degenerate and chaos must be detected by means other than level statistics \cite{Chen:2024oqv,Chen:2026vml,Belin:2026vfd}. In contrast, our focus will be on the non-BPS spectrum: away from the solvable points very little is known about the non-BPS states beyond perturbation theory (see however \cite{Afkhami-Jeddi:2021qkf,Keller:2023ssv}). We simply ask: \emph{are the non-BPS scaling dimensions Wigner-Dyson distributed as we deform away from a solvable point?} The natural data to analyze is the spectrum of primaries at fixed spin, which removes the obvious spectral correlations due to Virasoro symmetry and modular $T$-invariance. Any additional symmetries have to be resolved as well, since symmetries superpose statistically independent spectra and dilute their correlations \cite{Guhr:1997ve}.

The mechanism we use was identified in \cite{Srdinsek:2020bpq} for the double sine-Gordon model: the exactly degenerate spectrum of the free theory is lifted at first order in perturbation theory by a matrix whose eigenvalue spacings are Wigner-Dyson distributed when the two cosines are mixed with comparable coefficients, crossing over to Poisson statistics near the integrable limits. At first order the overall coupling only rescales this matrix, so the statistics depend only on the mixing. In that model the perturbations are relevant, so the deformed theory is a massive QFT and the first-order levels are finite-volume energies. We realize the same mechanism with an exactly marginal deformation, along which the theory remains conformal, so that the first-order splitting is a spectrum of anomalous dimensions of two-dimensional CFTs.

The same mechanism can be observed in perturbative anomalous dimensions \cite{McLoughlin:2020zew,McLoughlin:2022jyt,Baume:2026jyt} (see also \cite{Caputa:2026hvh}), in the matrices of second-order anomalous dimensions of $\frac14$-BPS states of the deformed symmetric orbifold of $\Tf$ at finite $N$ \cite{Zhang:2026jnf,Gaberdiel:2026jor}, and in spin systems with highly degenerate spectra \cite{Abdelshafy:2025ddx}. By contrast, in the planar limit of the symmetric orbifold the singlet and triplet perturbations admit an asymptotic integrable description \cite{Gaberdiel:2023lco,Frolov:2023pjw,Gaberdiel:2025smz} (see however \cite{Fabri:2025rok}), while the second-order shifts at fixed energy are close to Poisson \cite{Zhang:2026jnf}.\footnote{The planar limit suppresses the mixing of different cycle structures that leads to level repulsion at finite $N$. The Kummer model is closely related to $N=2$.}

At the Kummer point of the square torus at the free-fermion radius, the scalar states in the symmetry sector studied below form highly degenerate blocks at integer energies. The first-order splitting of each degenerate block by an exactly marginal deformation is then given by a finite matrix that can be computed using only free-field data. Figure~\ref{fig:spectrum} shows the spectrum and its deformation for the first six blocks. The moduli that blow up the $16$ fixed points symmetrically span a four-dimensional space of exactly marginal deformations, whose directions form a three-sphere of unit quaternions. Along a {\it generic} direction in this space, we find that the first-order shifts in fully resolved symmetry sectors exhibit GOE spectral statistics. Genericity is essential: we identify special directions on the three-sphere (among them the singlet deformation studied throughout the D1-D5 literature), which lead to deformations with enhanced symmetry either of the spectral sector we study or of the full theory.

\section{The K3 model and the space of deformations}
\label{sec:theory}

\paragraph*{The Kummer point.}
We start from the free $c=6$, $\mathcal N=(4,4)$ theory of four real bosons $X_{A\dot A}$ (the coordinates of a square torus $\Tf$), plus four left- and four right-moving real fermions $\psi^{\alpha A}$ and $\bar\psi^{\dot\alpha A}$, in the conventions of \cite{Guo:2019ady} (see also \cite{Avery:2010er}). The indices $A,\dot A$ are spinor indices of the $SU(2)_1\times SU(2)_2$ rotations of the tangent space of $\Tf$, an ``internal'' symmetry of the free fields that is broken by the lattice of the torus. Similarly, $\alpha,\dot\alpha$ are spinor indices of the two R-symmetries, $SU(2)_L\times SU(2)_R$. We work in the Ramond sector, in units $\alpha'=2$, with all four circles at the free-fermion radius $R=2$, where each compact boson is equivalent to a free Dirac fermion \cite{Ginsparg:1988ui}. The Fock states are labeled by their left- and right-moving momenta and integer oscillator levels (Supplemental Material (SM)~\cite{SupplementalMaterial}, Sec.~\ref{SM-app:kummer}); weights are measured relative to the Ramond ground state energy $\frac{c}{24}=\frac14$.

The Kummer orbifold gauges the reflection $(X,\psi,\bar\psi)\to-(X,\psi,\bar\psi)$ and gives the K3 sigma model at an orbifold point of its moduli space \cite{Dijkgraaf:1996xw,Seiberg:1999xz}. Its untwisted sector consists of the reflection-even states. In addition, each of the $16$ fixed points $x=\pi R\,\bff$, $\bff\in\{0,1\}^4$, carries a twisted sector in which all fields are antiperiodic. The ground state energies of the twisted bosons and fermions cancel, and the twisted states have the weights:
\begin{equation}
\label{eq:weights_tw}
h=N^{\rm tw}_L\,,\qquad \bar h=N^{\rm tw}_R\,,\qquad N^{\rm tw}_{L,R}\in\tfrac12\mathbb Z_{\ge0}\,,
\end{equation}
with $N^{\rm tw}_L+N^{\rm tw}_R\in\mathbb Z$ imposed by the orbifold projection (SM Sec.~\ref{SM-app:kummer}). 

Below, we will evaluate matrix elements of twist deformations of the Kummer model. To this end, it is useful to note that the model is closely related to the symmetric orbifold $\Sym^2(\Tf)$ \cite{Lunin:2001pw,Avery:2010er,Guo:2019ady} and to employ the associated covering space methods. The correspondence and momentum conventions are reviewed in SM Sec.~\ref{SM-app:firstorder}.\\

\paragraph*{Symmetry sectors.}
The half-period translations $U_\bt\colon  x\mapsto x+\pi R\,\bt$, $\bt\in\{0,1\}^4$, commute with the reflection and form a $\mathbb Z_2^4$ symmetry of the orbifold \cite{Taormina:2011rr,Gaberdiel:2013psa}, which is preserved by the deformations considered below. We label its $16$ eigenspaces, the symmetry sectors, by $\bs\in\{0,1\}^4$:
\begin{equation}
 {\cal H}_\bs = \big\{ |\Psi\rangle\colon \;\; U_\bt |\Psi \rangle = (-1)^{\bs \cdot \bt} |\Psi \rangle  \text{ for all } \bt \big\}\,.
\end{equation}
Such a sector contains the untwisted states whose momentum quantum numbers are congruent to $\bs$ modulo $2$ and one copy of the twisted tower \eqref{eq:weights_tw}, built by acting with oscillator modes on the combination $\frac14\sum_\bff(-1)^{\bs\cdot\bff}|\bff\rangle$ of the twisted ground states.\footnote{Note that $U_\bt |\bff\rangle = |\bff+\bt\rangle$, so $\frac14\sum_\bff(-1)^{\bs\cdot\bff}|\bff\rangle$ diagonalizes $U_{\bt}$.} At spin zero, the untwisted energies of a sector $\bs$ lie on $\mathbb Z+\frac14|\bs|$, where the weight $|\bs|$ is the number of non-zero components of $\bs$, while the twisted energies are integers. Since the deformation introduced below is built on a twist field, it only connects untwisted and twisted states, so that first-order mixing can only occur in the invariant sector $\bs=0$ and at the odd levels of the sector of weight $|\bs|=4$ (SM Sec.~\ref{SM-app:kummer}). We focus on the invariant sector.

In the invariant sector, reversing the right-moving fields alone, $(\bar\partial X,\bar\psi)\to-(\bar\partial X,\bar\psi)$, is a further involution, which is a symmetry of this sector only (SM Sec.~\ref{SM-app:kummer}). The deformations discussed below preserve it and we refer to it as $Z_R$.\\

\paragraph*{The marginal deformations.}
For each fixed point $\bff$, the corresponding twisted sector contains an NS-NS chiral primary $\twist_\bff^{++}$ of weights $h=\bar h = \frac12$ and R-charges $J_0^3 = \bar{J}_0^3=\frac12$. Acting on it with the supercurrents $G^\alpha_{\dot A}=\psi^{\alpha A}\partial X_{A\dot A}$ and $\bar G^{\dot\alpha}_{\dot A}=\bar\psi^{\dot\alpha A}\bar\partial X_{A\dot A}$, we generate the superdescendants $G^{-}_{\dot A,-\frac12}\bar G^{-}_{\dot B,-\frac12} \twist_{\bff}^{++}$. These have weights $h=\bar h=1$ and generate four real exactly marginal deformations associated with $\bff$, often called blow-up modes \cite{Dixon:1986qv,Nahm:1999ps}. We consider the four translation-invariant combinations of
these deformations, with equal coefficients at all fixed points:
\begin{equation}
\label{eq:Dc}
\Dc=\frac{1}{2\sqrt2}\,\cMat^{\dot A\dot B}\,G^{-}_{\dot A,-\frac12}\bar G^{-}_{\dot B,-\frac12}\twist^{++} \; + \; \text{h.c.}\,,
\end{equation}
where $\twist^{++}=\frac14\sum_\bff\twist^{++}_\bff$.
The Hamiltonian on the circle is deformed at first order:
\begin{equation}
\label{eq:HDc}
H(\lam)=H_0+\lam\,\Dcz \,,\quad\;\; \Dcz=\int_0^{2\pi}d\sigma\,\Dc(\sigma,0)\,.
\end{equation}
Each $\Dc$ is the top component of a short multiplet and hence an exactly marginal deformation that preserves conformality and $\mathcal N=(4,4)$ supersymmetry. Hermiticity of $\Dc$ implies that only the combination $\frac{1}{2}(\cMat+\epsilon \cMat^* \epsilon^T)$ contributes to the coefficient matrix. We can therefore take $\cMat^{\dot A \dot B}$ to be a real quaternion:
\begin{equation}
\label{eq:quaternion}
\cMat=q_0 \, \mathds{1}+ i \,(q_1\sigma_x+q_2\sigma_y+q_3\sigma_z)\,,\qquad q\in\mathbb R^4\,.
\end{equation}
The normalization $\langle\Dc(z)\Dc(0)\rangle=|q|^2/|z|^4$ fixes $|q|=1$, so the directions of deformation form a three-sphere $q\in S^3$. The unit $q=(0,0,-1,0)$, for which $\cMat^{\dot A\dot B}=\epsilon^{\dot A\dot B}$, is the $SU(2)_2$-invariant singlet deformation that is standard in the D1-D5 literature \cite{Avery:2010er,Guo:2019ady}. The other three units span the triplet of \cite{Gaberdiel:2025smz} (see also \cite{Keller:2019suk} and SM Sec.~\ref{SM-app:deformations}).\\

\paragraph*{States.}
We restrict to the Ramond-Ramond sector (the other sectors follow by spectral flow). For each copy of a massive representation of the two $\mathcal N=4$ algebras, we keep the superconformal primary with $J^3_0=\bar J^3_0=-\frac12$. In a massive representation this is the lowest weight of the R-spin-$\frac12$ doublet \cite{Eguchi:1987wf}. We consider scalar primaries, $h=\bar h$, and denote by $E=h+\bar h\ge0$ the energy above the Ramond-Ramond ground state. At $E=0$ the only state with these properties is the protected Ramond vacuum, the lowest weight of the R-spin-$(\frac12,\frac12)$ ground states. At the undeformed orbifold point, the selected primaries at each integer energy $E>0$ span a highly degenerate subspace $\B(E)$, which we call a {\it block} (see Fig.~\ref{fig:spectrum} for the first six block sizes).\\

\paragraph*{Symmetry resolution.}
The deformation \eqref{eq:HDc} preserves the two $\mathcal N=4$ algebras and breaks the additional continuous internal symmetries of the orbifold point (see SM Sec.~\ref{SM-app:symmetries}). Having removed the superconformal multiplet degeneracies by our choice of primaries, we now need to resolve the remaining discrete symmetries within each block ${\cal B}(E)$.

Discrete symmetries organize the space of directions $q\in S^3$. The point group of the square torus, worldsheet parity $\cMat\mapsto \cMat^T$, $\cMat\mapsto-\cMat$ and, on the invariant sector, a rotation of the right-movers alone all relate different directions with identical spectra in the invariant sector. They act on $S^3$ as reflections in $16$ mirrors, i.e., the great two-spheres $q_\mu=0$ and $q_\mu=\pm q_\nu$ with $\mu\neq\nu$ (End Matter). On a mirror, the corresponding reflection becomes an additional symmetry, which is reflected in the level statistics.

The singlet lies at a locus where nine mirrors meet, and the triplet spans the mirror $q_2=0$. Figure~\ref{fig:spectrum} (right) shows the effect on the first-order shifts of one sub-block along a great circle from the singlet through the reference direction to the mirror $q_2=0$: at the singlet the 25 shifts are highly degenerate, and away from it they split and repel. 

Away from the mirrors, the point-group transformations that leave $\cMat$ unchanged form the stabilizer $\Hc$. On the scalar primaries considered here, its action reduces to two commuting $\mathbb Z_2$ symmetries, whose eigenvalue signs label the four one-dimensional characters of $\Hc$ (the trivial character has both signs positive). The commuting involution $Z_R$ provides a third sign, giving eight possible sub-blocks at each level. Since all twisted scalar states at level $E$ have $Z_R=(-1)^E$, the four sub-blocks with $Z_R=-(-1)^E$ contain only untwisted states, which do not mix at first order and stay at $E$. At the even levels there is one further, hidden $\mathbb Z_2$, which we found numerically (End Matter). We present most of our numerics for a reference direction on $S^3$, drawn at random, which lies far from every mirror ($5.5^\circ$ from the nearest one):
\begin{equation}
\label{eq:qgen}
q_{\rm ref}=(-0.318,\ 0.096,\ -0.760,\ 0.559)\,.
\end{equation}

\section{First-order degenerate perturbation theory}
\label{sec:firstorder}

\begin{figure*}[t]
\centering
\includegraphics[width=\textwidth]{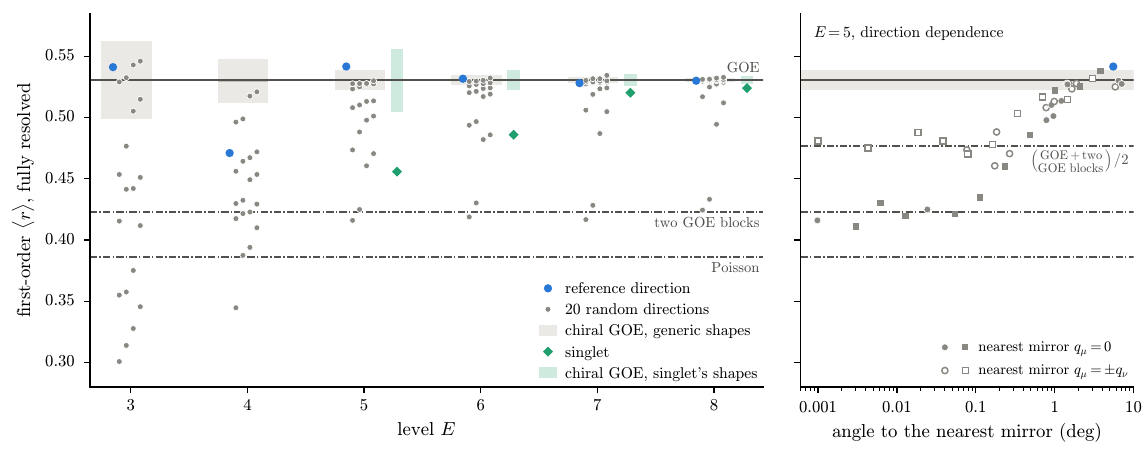}
\caption{{\it Left:} Symmetry-resolved first-order gap ratio along the reference direction (blue), along $20$ further directions drawn at random (gray), and along the singlet direction (green). Gray and green bands show the mean and standard deviation of $\rmean$ for chiral GOE matrices with shapes matched to the numbers of twisted states and distinct positive singular values. The singlet points resolve the singlet's own symmetries, including the hidden $\mathbb Z_2$ at the even levels. {\it Right:} $\rmean$ at $E=5$ for the reference direction plus the same $20$ random directions (circles) and $20$ further directions close to a single mirror (squares). We show the dependence of $\rmean$ on the angle $\delta$ to the nearest mirror, which illustrates the signatures of split sub-blocks in the vicinity of different types of mirrors ($\delta\to 0$).}
\label{fig:firstorderE}
\end{figure*}

\paragraph*{The first-order matrix.}
Let $\{|\Psi_a\rangle\}_{a=1}^{d_E}$ be an orthonormal basis of the degenerate block $\B(E)$. The first-order energy shifts are $\lam$ times the eigenvalues of $\langle\Psi_b|\Dcz|\Psi_a\rangle$. The level-preserving part of $\Dcz$ commutes with both $\mathcal N=4$ algebras, so the problem closes on superconformal primaries. Further, since the deformation contains a twist field, it only couples the $d_+$ untwisted states to the $d_-$ twisted states, with $d_E=d_++d_-$. In the basis of untwisted followed by twisted states, the first-order matrix is
\begin{equation}
\label{eq:Mc}
\HB^{(1)}=\begin{pmatrix}0&\Mc^{\dagger}\\ \Mc&0\end{pmatrix}\,.
\end{equation}
The $d_-\times d_+$ matrix $\Mc$ is constructed from free-field chiral twist amplitudes, as explained in End Matter. If $\nu_i$ are its non-zero singular values, the levels split as $E\pm\lam\nu_i$, while $d_++d_--2\rank\Mc$ states remain at $E$.\\

\paragraph*{Exact structures in the perturbed spectrum.}
The orbifold quantum $\mathbb Z_2$, $\mathcal Q$, acts as $+1$ on untwisted and $-1$ on twisted states \cite{Gaberdiel:2013psa}. It anticommutes with $\Dcz$, giving the paired shifts $\pm\nu_i$. The $Z_R$ selection rule above bounds the rank of $\Mc$ by the number of untwisted states with $Z_R=(-1)^E$ (End Matter), so that roughly half of the states of each level remain at $E$ at first order. In all generic cases we checked, the non-zero singular values are distinct; the statistics below concern these values.\\

\paragraph*{Gap ratios.}
For a generic direction a fully resolved sub-block is one of the four characters of $\Hc$ with $Z_R=(-1)^E$, split further by the hidden $\mathbb Z_2$ at the even levels. In each of them we sort the distinct singular values, $\nu_1<\dots<\nu_n$, and form the gap ratios:
\begin{equation}
\label{eq:r}
r_i=\frac{\min(\delta\nu_i,\delta\nu_{i+1})}{\max(\delta\nu_i,\delta\nu_{i+1})}\,,\qquad \delta\nu_i=\nu_{i+1}-\nu_i\,.
\end{equation}
We then combine (``pool'') the ratios of all sub-blocks of a level into a single sample and compute their mean $\rmean$ \cite{Oganesyan:2007wpd,Atas:2013gvn}. Its reference values are $0.5307$ for the GOE, $0.5996$ for the Gaussian unitary ensemble (GUE), $0.3863$ for Poisson statistics, and $0.423$ for a superposition of two independent GOE spectra of equal size, the signature of an unresolved $\mathbb Z_2$ \cite{Giraud:2020mmb}. We compare with the chiral GOE (class BDI in the Altland-Zirnbauer classification \cite{Verbaarschot:1994qf,Altland:1997zz}; see End Matter), i.e., matrices of the form \eqref{eq:Mc} with $\Mc$ replaced by a real Gaussian random matrix of the same shape.\\

\paragraph*{Results.}
In Fig.~\ref{fig:firstorderE} we show the gap ratios for the reference direction and other random directions, as well as for the singlet (see the table in End Matter for the underlying data). At every level except $E=4$ the reference direction agrees with the GOE value to within $1.4\,\sigma$ (for $E=4$, where the deficit is common to all directions, see End Matter). Resolving every symmetry, including the hidden $\mathbb{Z}_2$, is essential: at $E=6$, before the hidden $\mathbb Z_2$ is resolved, the reference direction gives $\rmean_{\rm ref}=0.509$, well below the GOE value. At $E=8$, the sub-block transforming trivially under the stabilizer gives $\rmean_{\rm ref}=0.434$ before this $\mathbb Z_2$ is resolved, and $0.536$ and $0.531$ in its two resolved eigenspaces (End Matter).

The right panel of Fig.~\ref{fig:firstorderE} shows a sample of random directions for $E=5$: directions that lie farther from the mirrors agree better with GOE statistics. Closer to a mirror, the additional symmetries are only weakly broken, which washes out the RMT statistics.  On a mirror $q_\mu=0$ these symmetries exactly split every sub-block, giving the value of two independent spectra, while on a mirror $q_\mu=\pm q_\nu$ they map two characters of $\Hc$ onto each other and split the others. As the level spacing decreases with $E$, the distance from a mirror required to break its symmetry also shrinks.

The singlet direction is fixed by the whole point group, and the hidden $\mathbb Z_2$ splits several of its sub-blocks at the even levels (End Matter). Even fully resolved, its gap ratio lies below the GOE value at every level, well below at $E=5,6$ and by about two standard deviations at $E=7,8$. Unlike a generic direction, it also features exact degeneracies within its resolved sub-blocks; e.g., at $E=8$, $583$ out of $10721$ non-zero values repeat, most of them at a single value $\nu=\pi$.\\


\paragraph*{Wigner surmise and number variance.}
The gap ratio tests the correlation of neighboring levels without reference to the mean level density. To test correlations over many levels we unfold each sub-block separately with a smooth fit of its counting function $N(\nu)$, the number of levels not larger than $\nu$, and then compute the distribution $P(s)$ of the unfolded nearest-neighbor spacings and the number variance $\Sigma^2(L)$, i.e., the variance of the number of unfolded levels in an interval of length $L$ (see SM Sec.~\ref{SM-app:firstorder}). Figure~\ref{fig:spacings} shows that at $E=8$ the spacing distribution follows the GOE Wigner surmise $P(s)=\frac{\pi s}{2}e^{-\pi s^2/4}$ over the sampled range, and that the number variance follows the GOE prediction \cite{Mehta2004} up to $L=10$ at $E=7,8$. At $E=6$ it rises above the GOE curve for $L\gtrsim4$, but the sub-blocks of this level are too small for a reliable separation of the fluctuations from the mean level density (SM Sec.~\ref{SM-app:firstorder}).

We again contrast with the unfolded spectrum of the singlet deformation, which matches the GOE surmise less well for nearest-neighbor spacings and deviates significantly from the GOE in the number variance (Fig.~\ref{fig:spacings}). The reason is that the singlet's first-order levels cluster at special values: neighboring levels still repel, which keeps its gap ratio relatively close to the GOE value at $E=7,8$, but the spectrum lacks long-range rigidity.

\ifdefined\arXivversion
  \begin{figure*}[t] 
  \centering
  \includegraphics[width=.85\textwidth]{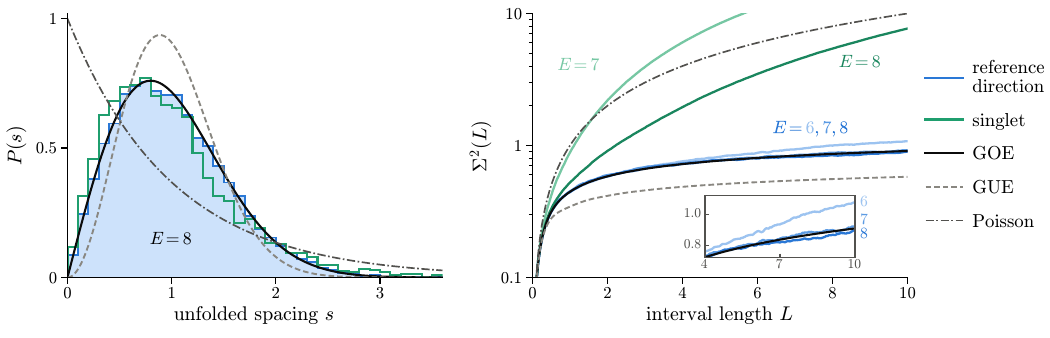}
\else
  \begin{figure}[b] 
  \centering
  \includegraphics[width=\columnwidth]{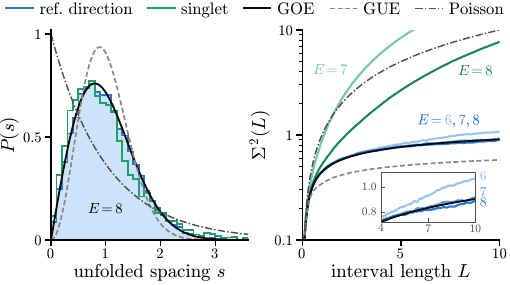}
\fi
\caption{Unfolded first-order spectra along the reference direction (blue) and the singlet (green). {\it Left:} Distribution of the nearest-neighbor spacings at $E=8$, compared with Wigner surmises and the Poisson distribution.
{\it Right:} The number variance measuring long-range correlations.}
\label{fig:spacings}
\ifdefined\arXivversion\end{figure*}\else\end{figure}\fi

\section{Conclusions}
\label{sec:conclusion}

We have shown that, deformed away from the Kummer orbifold along a generic direction, the K3 sigma model has primaries whose leading anomalous dimensions exhibit the level statistics of the GOE. The statistics are computed from the first-order matrices of an exactly marginal deformation inside the degenerate blocks of the free orbifold. As standard degenerate perturbation theory dictates, these matrices are computed from free-field data, so the statement holds for the deformed theory as $\lam\to0^+$. At finite coupling it holds only while the second-order shifts stay small compared to the first-order spacings.\footnote{Random matrix level statistics at weak coupling does not by itself imply ergodicity (cf.\ \cite{Abdelshafy:2025ddx}): $H_0$ is conserved to this order, and different levels remain uncorrelated.} The result requires a generic direction on the quaternion sphere of $\mathbb Z_2^4$-symmetric blow-up moduli and the resolution of every exact symmetry. Special directions, among them the well-known singlet deformation, carry additional symmetries or extra structures that wash out random matrix statistics. We identified the hidden $\mathbb Z_2$ only numerically and it would be interesting to understand its origin.

Our analysis is restricted to the invariant sector $\bs=0$. The odd levels of the sector of weight $|\bs|=4$ provide a separate first-order problem, while the remaining $14$ sectors are split only at second order. Using the relation to the symmetric orbifold $\Sym^2(\Tf)$, it would be interesting to follow the first-order statistics in $\Sym^N(\Tf)$ as $N$ grows, where the planar second-order shifts are close to Poisson \cite{Zhang:2026jnf,Gaberdiel:2026jor}, and to connect the onset of random matrix statistics at infinitesimal coupling with the random matrix universality expected for holographic CFTs. It would also be interesting to relate the random matrix statistics of the non-BPS spectrum found here to the proposed diagnostics of BPS chaos \cite{Chen:2024oqv,Chen:2026vml,Belin:2026vfd}.

Beyond first order the spectrum can be computed using an energy truncation $E\le\Lam$, renormalized in the manner of Hamiltonian truncation \cite{Hogervorst:2014rta,Rychkov:2014eea,Elias-Miro:2017xxf}: at second order, the counterterms that remove its cutoff dependence follow from the operator product of the perturbation with itself. The analysis of the truncation and the second-order primary spectrum will be presented elsewhere \cite{toappear}. Exact marginality suggests a second route: at every coupling the deformed theory is again a CFT, so the level dynamics becomes a flow of the conformal data along the conformal manifold \cite{Pechukas:1983,Yukawa:1985zz}, which can in principle be integrated \cite{Behan:2017mwi}. As the coupling grows, the first-order bands of neighboring levels touch and their spectra merge. It is worth investigating whether random matrix statistics then extend across levels, and to look for other signatures of chaos, such as eigenstate thermalization \cite{Deutsch:1991msp,Srednicki:1994mfb,Lashkari:2016vgj}.\\


\paragraph*{Acknowledgments.}
We thank A.\ Dei and I.\ Zadeh for discussions and comments. FMH is supported by UK Research and Innovation (UKRI) under the UK government's Horizon Europe funding Guarantee EP/X030334/1. MR is supported by an NSERC Discovery Grant and used the clusters of the Digital Research Alliance of Canada for the numerical computations. AI tools (Claude Opus and Fable, ChatGPT Pro for Academic Researchers) were used for the numerical implementation and plots, for some analytic calculations and for refining ideas.

\paragraph*{Data availability.}
The code and data that support the findings of this Letter are openly available~\cite{HaehlRozali:2026code}.

\nocite{Larsen:1999uk,Keller:2011xi,Hartman:2014oaa,Gava:2002xb,Gaberdiel:2015uca,Keller:2019yrr,Guo:2020gxm}
\bibliographystyle{apsrev}
\bibliography{references}


\appendix
\section*{End Matter}
\setcounter{equation}{0}
\renewcommand{\theequation}{A\arabic{equation}}

\paragraph*{Matrix elements and rank bound.}
The states of a degenerate block $\B(E)$ are linear combinations of two kinds of Fock states (SM Sec.~\ref{SM-app:kummer}): twisted states $|t\rangle=|t^Lt^R\rangle$ are products of chiral oscillator states on the invariant twisted ground state, while untwisted states $|u\rangle$ are reflection-even combinations of products $|u^Lu^R\rangle$. The spatial integral over the deformation operator conserves the spin, and between spin-zero states at the same level it gives $\langle\Psi_b|\Dcz|\Psi_a\rangle=2\pi\langle\Psi_b|\Dc(0,0)|\Psi_a\rangle$. Finally, the level-preserving part of $\Dcz$ maps primaries to primaries because it commutes with the two $\mathcal N=4$ algebras (SM Sec.~\ref{SM-app:firstorder}).

Between products of chiral states, the local matrix elements factorize:
\begin{equation}
\label{eq:factor}
\begin{aligned}
&\big\langle t_b\big|G^{-}_{\dot A,-\frac12}
 \bar G^{-}_{\dot B,-\frac12}\twist^{++}
 \big|u_a\big\rangle=\langle t_b^L|{\cal A}_{\dot A}|u_a^L\rangle
         \langle t_b^R|{\cal A}_{\dot B}|u_a^R\rangle\,.
\end{aligned}
\end{equation}
Matrix elements between reflection-even or symmetry-adapted states follow by linearity. The chiral twist maps ${\cal A}_{\dot A}$ are computed with the covering space method of the symmetric orbifold following \cite{Lunin:2001pw,Avery:2010er,Guo:2019ady}. Only their level-preserving parts enter above. With the momentum-lattice and orbifold restrictions understood, the rectangular matrix in the first-order Hamiltonian \eqref{eq:Mc} is
\begin{equation}
\label{eq:Mc-chiral}
\Mc=\sqrt2\,\pi\,\cMat^{\dot A\dot B}
{\cal A}_{\dot A}\otimes{\cal A}_{\dot B}\,.
\end{equation}
At each level, the chiral maps are finite matrices and the construction is algorithmic.

The grading by $\mathcal Q$ guarantees at least $|d_+-d_-|$ zero modes. A stronger bound follows from $Z_R$. At level $E$, $Z_R$ reverses every right-moving mode, so that all twisted scalar states have $Z_R=(-1)^{2\bar h}=(-1)^E$ and only untwisted states with this eigenvalue can mix with them. If their number is $\tilde d_+$, then
\begin{equation}
\label{eq:rankbound}
\rank(\Mc)\le\tilde d_+\,.
\end{equation}
At odd levels, $Z_R$ pairs the untwisted momentum orbits without fixed points, so $\tilde d_+=d_+/2$ (SM Sec.~\ref{SM-app:kummer}). At even levels its two eigenspaces need not have equal dimensions, and we compute $\tilde d_+$ directly. Along a generic direction the bound is saturated at $E=3,5,7$; at $E=4,6,8$ saturation fails by $1$, $84$, $57$ states, respectively. Thus the odd-level zero-mode count is as small as allowed by the grading and the rank bound. The total number of unshifted states is $d_++d_--2\rank\Mc$.\\

\paragraph*{Symmetry resolution.}
The point group of the square torus ($192$ orientation-preserving signed permutations of the coordinates), worldsheet parity $\cMat\mapsto \cMat^T$, and $\cMat\mapsto-\cMat$ relate equivalent deformed theories. Their action forms a reflection group whose ten mirrors are the intersections of $S^3$ with the hyperplanes $q_2=0$, $q_\mu=0$ and $q_\mu=\pm q_{\nu\neq\mu}$ with $\mu,\nu\in\{0,1,3\}$. Each mirror is a great two-sphere on which the corresponding reflection becomes an additional symmetry of the deformed theory. The invariant sector has one further equivalence: a rotation of the right-movers alone, which maps the deformation along $\cMat$ onto that along $\cMat\,i\sigma_x$, $\cMat\,i\sigma_y$ or $\cMat\,i\sigma_z$, so that their invariant-sector spectra coincide exactly (SM Sec.~\ref{SM-app:symmetries}). This adds the six mirrors $q_2=\pm q_\mu$, on which the corresponding reflection is an additional symmetry of the invariant sector. The $16$ mirrors cut $S^3$ into $384$ spherical tetrahedra, and no two directions in the same tetrahedron are related by the above discrete maps.\\

The symmetries we can easily identify along a generic direction, i.e., the stabilizer $\Hc$ and $Z_R$, are not enough to resolve every sub-block. For even levels, the gap ratio of the sub-block of the trivial character of $\Hc$ with $Z_R=+1$ is that of two superposed spectra, the signature of an unresolved symmetry~\cite{Giraud:2020mmb}. 

An exact symmetry commutes with $H_0$ and with $\Dcz$, including its matrix elements between different levels, which first-order perturbation theory does not use. The first-order matrix of one level is therefore not enough to identify such a symmetry, since operators built from its own singular vectors commute with it as well. In practice, we thus look numerically for subspaces that $H_0$ and $\Dcz$ leave jointly invariant, using the matrices projected onto the selected primaries through $E=6$, including off-level matrix elements up to $E=6$ together with the matrix elements of $\Dcz$ between different levels. A symmetry commuting with both acts within each level and would be diagonal in the eigenbasis of $H(\lambda)=H_0+\lambda\Dcz$ for every $\lambda$. Therefore, eigenvectors computed at two different couplings (we choose $\lambda=0.2$ and $0.35$) are orthogonal when they lie in different eigenspaces of the symmetry, while those in the same eigenspace generically overlap. The eigenspaces can thus be found by computing all eigenvectors at each of the two couplings and linking an eigenvector at one coupling to one at the other whenever their overlap exceeds a small threshold. The connected components of the resulting graph are the eigenspaces. 

\begin{figure}[t]
\centering
\includegraphics[width=\columnwidth]{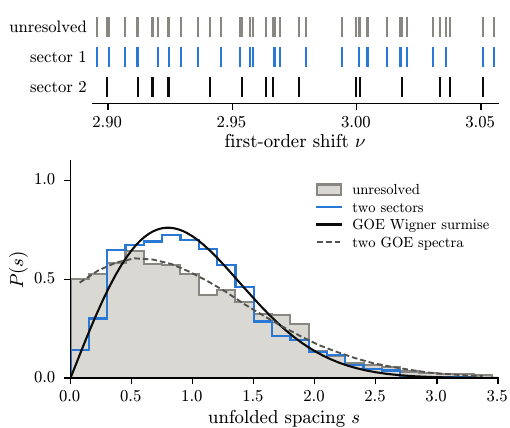}
\caption{The hidden $\mathbb Z_2$ along the reference direction for $E=6$, in the sub-block of the trivial character of $\Hc$ with $Z_R=+1$: a total of $1350$ first-order values split into eigenspaces of $899$ and $451$ values. {\it Top:} $40$ consecutive levels of the unresolved sub-block and the same levels split into the two eigenspaces of the $\mathbb Z_2$. {\it Bottom:} Distribution of the unfolded nearest-neighbor spacings for the unresolved sub-block (gray) and for its two eigenspaces (blue), compared with the GOE Wigner surmise and two independent GOE spectra superposed in the proportion $899:451$ (dashed).}
\label{fig:hiddenZ2}
\end{figure}

An unresolved symmetry superposes two independent spectra, whose levels do not repel each other. Figure~\ref{fig:hiddenZ2} shows this for the reference direction at $E=6$: in the $40$ levels at the top, levels that come very close almost always lie in different eigenspaces. The distribution of unfolded spacings (bottom) shows the same: for the unresolved sub-block, $P(s)$ stays finite as $s\to0$, like that of two independent GOE spectra superposed in the proportion $899:451$, whereas within each eigenspace the levels repel as in the GOE. The gap ratio confirms this: the unresolved sub-block has $\rmean_{\rm ref}=0.432$, close to the value $0.441\pm0.009$ of the two superposed GOE spectra. Its two eigenspaces give $0.524$ and $0.519$, respectively. Pooled with the three other sub-blocks of this level, they give the value $\rmean_{\rm ref} = 0.532$ reported in Sec.~\ref{sec:firstorder}. At $E=8$ we proceed differently. Since the involution does not depend on the direction (see below), it commutes with the first-order matrix of every direction, and the overlap test above can be applied to the singular vectors of $\Mc$ at two generic directions. At $E=4$ and $6$ this reproduces the eigenspaces found from the truncated Hamiltonian. At $E=8$ it splits the $9216$ values of the sub-block into eigenspaces of $6137$ and $3079$ values, with $\rmean_{\rm ref}=0.536$ and $0.531$.\\

The same $\mathbb Z_2$ acts along the singlet direction. The involution that splits the sub-block does not depend on the direction: it commutes with $\Dcz$ for every $\cMat$ and with the whole point group, but it is not an element of the point group. Along the singlet, whose larger symmetry group divides the states into smaller sub-blocks, it splits several of those with $Z_R=+1$ at the even levels, again with eigenspaces in a ratio of about two to one. It mixes oscillator states with momentum states of the same energy. Left unresolved, it lowers the singlet’s gap ratio from $0.486$ to $0.465$ at $E=6$ and from $0.524$ to $0.494$ at $E=8$.\\

\paragraph*{Statistical analysis.}
The chiral GOE is the expected symmetry class: CPT composed with the R-symmetry Weyl reflections defines an antiunitary symmetry $T$ with $T^2=+1$ that preserves each fully resolved sub-block along a generic direction. Its first-order matrix therefore admits a real symmetric representation (SM Sec.~\ref{SM-app:symmetries}). Moreover, the twisted states outnumber the untwisted states that couple to them by a factor of about two, so that the relevant matrices are far from square: the smallest singular value lies far from the hard edge at zero, and the bulk statistics are that of the GOE.

The following table gives the untwisted and twisted dimensions $d_\pm$, the rank of $\Mc$, the gap ratio for the reference direction (pooling fully resolved sub-blocks), and the standard deviation $\sigma$ of the mean gap ratio for a single chiral GOE spectrum of matrices with the same sizes:
\begin{equation*}
\begin{tabular}{cccccc}
\toprule
$E$ & $d_+$ & $d_-$ & $\rank \Mc$ & $\rmean_{\rm ref}$ & $\sigma$ \\
\midrule
3 & 200    & 196   & 100    & $0.541$ & $0.032$ \\
4 & 729    & 784   & 304   & $0.471$ & $0.018$ \\
5 & 2952   & 2916  & 1476  & $0.542$ & $0.0081$ \\
6 & 10420  & 9604  & 5448  & $0.532$ & $0.0042$ \\
7 & 28288  & 28224 & 14144 & $0.528$ & $0.0026$ \\
8 & \;76881\; & \;79524\; & \;36840\; & $ \;0.530\; $ & $\;0.0016\;$ \\
\bottomrule
\end{tabular}
\end{equation*}
To compute $\sigma$, we pool the gap ratios for $3000$ independent realizations, each containing one real Gaussian matrix per sub-block, with as many rows as twisted states and as many columns as distinct non-zero singular values. At $E=8$ the two-direction method does not determine how the twisted states divide between the two eigenspaces of the hidden $\mathbb Z_2$. We divide them in proportion to the sizes of the eigenspaces, and $\sigma$ does not depend on this choice at the quoted accuracy. We observe a significant deviation from the GOE expectation at $E=4$, which is a property of this level and not of the reference direction: over $3000$ random directions more than $4^\circ$ from every mirror, the pooled gap ratio is $0.495$ on average. In the three sub-blocks without a hidden $\mathbb Z_2$, the deficit comes from the upper end of the spectrum: the ten gap ratios at the top of each sub-block average $0.40$ and the remaining ones $0.53$. There the highest level is isolated and the two levels below it are nearly degenerate. Along the singlet direction the same sub-blocks have a nondegenerate top level and an eightfold degenerate one below it, which splits into two levels and three pairs at first order in the distance from the singlet. We found no comparable structure at $E=3,5,6$ and no further exact symmetry in these sub-blocks. The two eigenspaces of the hidden $\mathbb Z_2$ give $\rmean=0.45$ each.

\ifdefined\arXivversion
  \clearpage
  \onecolumngrid
  \SMsetup
  
\ifdefined\arXivversion
  \begin{center}\large\bfseries Supplemental Material\end{center}
  \setcounter{section}{0}
  \setcounter{table}{0}
\else
  \title{Supplemental Material: Onset of quantum chaos in a family of two-dimensional conformal field theories}
  \author{Felix M.\ Haehl}
  \author{Moshe Rozali}
  \maketitle
\fi
\setcounter{secnumdepth}{1}
\setcounter{equation}{0}
\setcounter{figure}{0}
\renewcommand{\thesection}{S\arabic{section}}
\renewcommand{\theequation}{S\arabic{equation}}
\renewcommand{\thefigure}{S\arabic{figure}}
\renewcommand{\thetable}{S\arabic{table}}
\renewcommand{\theHsection}{S\arabic{section}}
\renewcommand{\theHequation}{S\arabic{equation}}
\renewcommand{\theHfigure}{S\arabic{figure}}
\renewcommand{\theHtable}{S\arabic{table}}

\section{The Kummer point and its symmetry sectors}
\label{app:kummer}

\paragraph*{States on the torus.}
The free fields obey the reality conditions $(X_{A\dot A})^*=-X^{A\dot A}$, $(\psi^{\alpha A})^\dagger=-\psi_{\alpha A}$ and $(\bar\psi^{\dot\alpha A})^\dagger=-\bar\psi_{\dot\alpha A}$, with all indices raised and lowered by $\epsilon_{+-}=1=-\epsilon^{+-}$, so that $\psi^{\alpha A}$ and $\bar\psi^{\dot\alpha A}$ each contain two complex fermions \cite{Avery:2010er}. The four real components of $X_{A\dot A}$ can be traded for four bosonic coordinates along the four orthogonal circles of $\Tf$, $X^i$ $(i=1,\dots,4)$, each obeying $X^i\sim X^i+2\pi R$. On a single circle the states carry an integer momentum $n$ and an integer winding $w$. In units $\alpha'=2$, in which the self-dual radius is $\sqrt2$, their momenta and weights are:
\begin{equation}
\label{eq:circle}
p_{L,R}=\frac{n}{R}\pm\frac{wR}{2}\,,\qquad h=\frac12\,p_L^2+N_L\,,\qquad \bar h=\frac12\,p_R^2+N_R\,.
\end{equation}
At the free-fermion radius $R=2$, where each compact boson is equivalent to a free Dirac fermion, the momenta $a\equiv p_L=\frac n2+w$ and $b\equiv p_R=\frac n2-w$ are half-integers with $a-b=2w$ even. Conversely, every such pair $(a,b)$ belongs to a state with $n=a+b$ and $w=\frac12(a-b)$. Repeating this construction on all four circles of $\Tf$ gives the weights
\begin{equation}
\label{eq:weights}
h=\frac12|\ba|^2+N_L\,,\qquad \bar h=\frac12|\bb|^2+N_R\,,
\end{equation}
with oscillator levels $N_{L,R}\in\mathbb Z_{\ge0}$, left- and right-moving momenta $\ba,\bb\in(\frac12\mathbb Z)^4$ subject to $\ba-\bb\in2\mathbb Z^4$, integer momentum $\bn=\ba+\bb$ and winding $\bw=\frac12(\ba-\bb)$. The spin $h-\bar h=\frac12(\ba-\bb)\cdot(\ba+\bb)+N_L-N_R=\bn\cdot\bw+N_L-N_R$ is an integer, as it must be. Figure~\ref{fig:sectors} (left) shows the momenta $(a_i,b_i)$ of one circle.\\

\begin{figure}[htb]
\centering
\includegraphics[width=0.85\textwidth]{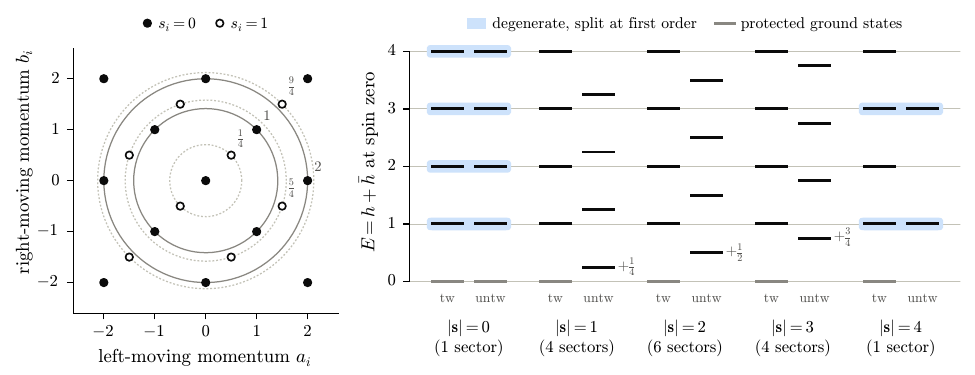}
\caption{The levels of the Kummer point by symmetry sector. {\it Left:} The left- and right-moving momenta $(a_i,b_i)$ of one circle of the torus: integer momenta ($s_i=0$) have integer energy $\tfrac12(a_i^2+b_i^2)$, half-integer ones ($s_i=1$) are shifted by $\tfrac14$. The circles are curves of constant energy $\tfrac12(a_i^2+b_i^2)$, labeled by its value (solid for $s_i=0$, dotted for $s_i=1$). {\it Right:} There are $16$ symmetry sectors labeled by $\bs \in \{0,1\}^4$. We show the twisted (tw) and untwisted (untw) energies of the spin-zero states in the sectors of weight $|\bs|$. The twisted tower \eqref{M-eq:weights_tw} is the same in every sector, while the untwisted energies lie on $\mathbb Z+\frac14 |\bs|$ (only on the odd integers for $|\bs|=4$). Twisted and untwisted states are degenerate only for $|\bs|=0$ and at the odd levels for $|\bs|=4$ (shaded).}
\label{fig:sectors}
\end{figure}

For each allowed momentum pair $(\ba,\bb)$, let $|\ba,\bb;0_R\rangle$ denote the Ramond oscillator vacuum of lowest left- and right-moving R-charge. We write the Fock-space states in the form:
\begin{equation}
|\ba,\bb;I_L,I_R\rangle_{\Tf} = {\cal O}_{I_L}\,\bar{\cal O}_{I_R}
|\ba,\bb;0_R\rangle\,,
\end{equation}
where $I_L,I_R$ specify the modes, polarizations and occupation numbers of the creation operators. The oscillator monomials are understood to be normalized and may include creation operators for fermion zero modes.\\

\paragraph*{The orbifold.}
The reflection maps a state of momenta $(\ba,\bb)$ to one of momenta $(-\ba,-\bb)$ and reverses the sign of every oscillator, including the fermion zero modes. The untwisted sector consists of the even combinations of a state and its image: each pair $(\ba,\bb)\neq(0,0)$ gives one reflection-invariant combination:
\begin{equation}
|\ba,\bb;I_L,I_R\rangle_{\text{untw}} \propto |\ba,\bb;I_L,I_R\rangle_{\Tf} + (-1)^{k_L+k_R} \, |-\ba,-\bb;I_L,I_R\rangle_{\Tf}\,,
\end{equation}
where $k_L$ ($k_R$) counts the creation operators in ${\cal O}_{I_L}$ ($\bar {\cal O}_{I_R}$). For $\ba=\bb=0$ the states with an even number of oscillator excitations $k_L+k_R$ survive as well. 

In the twisted sectors the fields are only periodic up to reflection, $X(\sigma+2\pi)=-X(\sigma)$. A twisted-sector string must be localized at a fixed point $x\equiv-x$ of the reflection, i.e., at one of the $16$ half-period points $x_\bff=\pi R\,\bff$, $\bff\in\{0,1\}^4$, so the twisted sector is a direct sum of $16$ fixed-point sectors (one for each conjugacy class of reflections in the space group of the torus). All its fields are antiperiodic, so their modes are half-integer and there are no momenta or windings. The zero-point energy of each of the four antiperiodic bosons lies $\frac1{16}$ higher than that of a periodic boson. Similarly, each of the four antiperiodic fermions lies $\frac1{16}$ below that of a Ramond fermion. The two cancel, so the ground state of every twisted sector has $h=\bar h=0$, exactly like the Ramond ground state of the untwisted sector. Every twisted mode is odd under the reflection and the twisted ground state is even, so a state built with $k$ modes has parity $(-1)^k$, while its total level $N^{\rm tw}_L+N^{\rm tw}_R$ lies in $\frac{1}{2}k+\mathbb Z$. The projection onto even states gives \eqref{M-eq:weights_tw}, and the tower is the same at every fixed point.\\

\paragraph*{Symmetry sectors.}
To organize the twisted states by the same quantum numbers as the untwisted states, we pass from the basis labeled by fixed points to the simultaneous eigenbasis of the half-period translations. The half-period translations $U_\bt\colon x \mapsto x+\pi R\, \bt$ commute with the orbifolding reflection, since $-(x+\pi R\,\bt)\equiv-x+\pi R\,\bt$ up to a lattice vector. The translations form a symmetry group $\mathbb Z_2^4$. We label its simultaneous eigenspaces by $\bs\in\{0,1\}^4$, specifying their joint eigenvalues:
\begin{equation}
 {\cal H}_\bs = \left\{ |\Psi\rangle: \quad U_\bt |\Psi\rangle = (-1)^{\bs \cdot \bt} |\Psi \rangle \text{ for every } \bt \right\} \,.
\end{equation} 
These symmetry sectors are not to be confused with the twisted and untwisted sectors, since each contains states of both kinds. On an untwisted state, $U_\bt$ acts by the phase $(-1)^{\bn\cdot\bt}$, so its symmetry label is $\bs=\bn\bmod2=(2\ba)\bmod2=(2\bb)\bmod2$. This means that in the sector $\bs$, the momenta $a_i$ and $b_i$ are integers for $s_i=0$ and half-integers for $s_i=1$. The twisted sectors are associated with the fixed points $x_{\bff} = \pi R \, \bff$ and thus they are permuted by the translations, $U_{\bt} \colon x_{\bff} \mapsto x_{\bff + \bt}$. The action of $U_\bt$ is diagonalized by the following combinations:
\begin{equation}
\label{eq:twcomb}
|\bs\rangle_{\rm tw}=\frac14\sum_{\bff\in\{0,1\}^4}(-1)^{\bs\cdot\bff}\,|\bff\rangle 
\qquad \Rightarrow \qquad U_\bt   |\bs\rangle_{\rm tw} = (-1)^{\bs\cdot\bt}\,|\bs\rangle_{\rm tw} \,,
\end{equation}
where $|\bff\rangle$ denotes the normalized twisted Ramond ground state at $x_\bff$. Since translations leave the oscillator modes unchanged, the same Fourier combinations organize every oscillator excitation into the corresponding symmetry sector. Each sector $\bs$ therefore contains the untwisted states with momentum $\bn\equiv\bs\bmod2$ and one copy of the twisted tower built on $|\bs\rangle_{\rm tw}$. The invariant sector $\bs=0$ contains the untwisted states of even momentum and the symmetric combination of the fixed-point sectors.

Permutations of the four circles are symmetries of the square torus, so two sectors whose labels differ by a permutation are isospectral. The $15$ sectors with $\bs\neq 0$ fall into four classes, labeled by the weight $|\bs|=1,2,3,4$ (the number of non-zero components of $\bs$), with $4$, $6$, $4$ and $1$ members. On a circle with $s_i=0$ the energy $\frac12(a_i^2+b_i^2)$ is an integer, while on a circle with $s_i=1$ it is an integer plus $\frac14$ (see Fig.~\ref{fig:sectors}). The untwisted energies of a sector $\bs$ therefore lie on $\mathbb Z+\frac14|\bs|$, whereas the twisted energies are integers in every sector. Twisted and untwisted states can thus be degenerate only in the invariant sector and in the sector of weight four. Since our marginal deformations only connect untwisted to twisted states, at first order they act only within blocks that contain states of both kinds. In the $14$ sectors of weight $|\bs|=1,2,3$ there are no such blocks: the first-order shifts vanish, and the degeneracies are lifted only at second order. In the sector of weight $|\bs|=4$ all momenta are half-integers, so $|\ba|^2$ is odd, and the energies $E=|\ba|^2+2N_L$ of its spin-zero untwisted states are odd integers. That sector therefore mixes twisted and untwisted states at first order only at its odd levels.\\

\paragraph*{The involution $Z_R$.}
The reversal of the right-moving fields alone, $(\bar\partial X,\bar\psi)\to-(\bar\partial X,\bar\psi)$, maps the momenta $(\ba,\bb)$ to $(\ba,-\bb)$:
\begin{equation}
 Z_R |\ba,\bb;I_L,I_R\rangle_{\text{untw}} = (-1)^{k_R} \, |\ba,-\bb;I_L,I_R \rangle_{\text{untw}} \,.
\end{equation}
This respects $\ba-\bb\in2\mathbb Z^4$ only if $\ba+\bb\in2\mathbb Z^4$, i.e., in the invariant sector $\bs=0$. Thus, $Z_R$ acts only in this sector. It commutes with the reflection, with the point group and with both $\mathcal N=4$ algebras, whose right-moving generators are even, so it maps the space of states to itself. On the oscillator vacua, we choose the conventional sign $Z_R |\ba,\bb;0_R\rangle = +|\ba,-\bb;0_R\rangle$.

First, consider the action of $Z_R$ on the twisted states. It reverses every right-moving mode, so with $Z_R=+1$ on the twisted ground state, a twisted state with $k_R$ right-moving modes has $\bar h\in\frac{1}{2}k_R+\mathbb Z$ and $Z_R=(-1)^{k_R}=(-1)^{2\bar h}$, which evaluates to $Z_R=(-1)^E$ for a scalar of level $E$.

Next, consider untwisted states. At an odd level, both $|\ba|^2=E-2N_L$ and $|\bb|^2=E-2N_R$ are odd, so neither $\ba$ nor $\bb$ vanishes. The momentum pairs $\{\pm(\ba,\bb)\}$ and $\{\pm(\ba,-\bb)\}$ are therefore disjoint and $Z_R$ maps the first pair to the second and vice versa. Its two eigenspaces at that level thus have equal dimensions. In particular, the number of untwisted primaries with $Z_R=(-1)^E$, which can couple to the twisted states, is $\tilde d_+=\frac12 d_+$.

Finally, we note that the marginal deformations preserve $Z_R$: indeed, as one can check explicitly for any oscillator excitation using the covering-space construction, the right-moving twist maps satisfy:
\begin{equation}
 \langle Z_R t_R|{\cal A}_{\dot B}|Z_R u_R\rangle
 =\langle t_R|{\cal A}_{\dot B}|u_R\rangle \,.
\end{equation}
Since the $\Dc$ matrix elements chirally factorize and $Z_R$ only acts on the right-moving half, it follows that $[Z_R,\Dc]=0$ on the invariant sector ${\cal H}_{\bs = 0}$.\\

\paragraph*{States used in the calculation.}
With the restriction to spin-zero Ramond states understood (Sec.~\ref{M-sec:theory}), we retain one primary from each copy of an irreducible massive representation of the two $\mathcal N=4$ algebras at $E>0$. In each chirality, its lowest-energy subspace contains an R-spin-$\frac12$ doublet and two singlets. We choose the lowest-weight component $|P\rangle$, characterized by:
\begin{equation}
\begin{split}
J^3_0|P\rangle &=-\tfrac12|P\rangle\,,\qquad G^-_{\dot A,0}|P\rangle=0\,,\\
L_n|P\rangle &=J^a_n|P\rangle =G^\alpha_{\dot A,n}|P\rangle=0 \qquad(n\geq1)\,,
\end{split}
\end{equation}
and similarly with bars. Applying these conditions to the orbifold Fock spaces, we construct complete orthonormal bases of $\B(E)$ for $0<E\le8$.

At $E=0$, our R-charge conditions select only the protected Ramond vacuum (Sec.~\ref{M-sec:theory}). For each level $E\leq 8$, we diagonalize the first-order deformation on the space of these primaries. States at other energies do not enter its first-order splitting.

\section{The marginal deformations}
\label{app:deformations}

\paragraph*{Supercurrents.}
In each chirality the free theory has a small $\mathcal N=4$ superconformal symmetry with $c=6$. In the conventions of \cite{Avery:2010er} the left-moving generators are the stress tensor, the $SU(2)_L$ R-currents and four supercurrents:
\begin{equation}
\label{eq:N4gen}
T=-\frac12(\partial X^\dagger)^{A\dot A}\partial X_{A\dot A}-\frac12(\psi^\dagger)_{\alpha A}\partial\psi^{\alpha A}\,,\qquad
J^a=-\frac14(\psi^\dagger)_{\alpha A}(\sigma^{Ta})^{\alpha}{}_{\beta}\,\psi^{\beta A}\,,\qquad
G^\alpha_{\dot A}=\psi^{\alpha A}\partial X_{A\dot A}\,,
\end{equation}
where $\sigma^a$ are the Pauli matrices. The right-moving generators are built in the same way from the right-moving fermions and the right-moving current of the same bosons. The supercurrents have $h=\frac32$. They are doublets of $SU(2)_L$, so that $G^{\pm}$ raise and lower the R-charge by $\frac12$, doublets of $SU(2)_2$, and singlets of $SU(2)_1$. Their modes obey $(G^\alpha_{\dot A,r})^\dagger=\epsilon_{\alpha\beta}\,\epsilon^{\dot A\dot B}\,G^{\beta}_{\dot B,-r}$, so they amount to two complex supercurrents \cite{Guo:2019ady}. The modes $G^\alpha_{\dot A,r}=\oint\frac{dz}{2\pi i }\,z^{r+\frac12}\,G^\alpha_{\dot A}(z)$ satisfy \cite{Avery:2010er}:
\begin{equation}
\label{eq:GG}
\{G^\alpha_{\dot A,m},G^\beta_{\dot B,n}\}=\epsilon_{\dot A\dot B}\left[\left(n^2-\frac14\right)\epsilon^{\alpha\beta}\delta_{m+n,0}
+(m-n)(\sigma^{aT})^{\alpha}{}_{\gamma}\,\epsilon^{\gamma\beta}J^a_{m+n}+\epsilon^{\alpha\beta}L_{m+n}\right]\,,
\end{equation}
where, as usual, $r\in \mathbb{Z}$ in the Ramond sector, while $r\in \frac12+\mathbb{Z}$ in the NS sector. A chiral primary is a superconformal primary that is annihilated by $G^{+}_{\dot A,-\frac12}$, and its weight equals its R-spin, $h=j$. Acting on a chiral primary with $h=j=\frac12$ on both sides, $G^{-}_{\dot A,-\frac12}\bar G^{-}_{\dot B,-\frac12}$ gives an R-singlet of weights $(1,1)$, the top component of a short multiplet. Integrated over the worldsheet, such an operator is an exactly marginal deformation that preserves $\mathcal N=(4,4)$ supersymmetry.\\

\paragraph*{The singlet and triplet deformations.}
In the D1-D5 literature the undeformed theory is the symmetric orbifold $\Sym^N(\Tf)$, and the deformation is built on the twist field $\twist^{\beta\dot\beta}$ of the sector in which two copies of the torus theory are exchanged \cite{Avery:2010er,Guo:2019ady}. It is a doublet of $SU(2)_L\times SU(2)_R$ with $h=\bar h=j=\bar j=\frac12$, whose highest R-weight component $\twist^{++}$ is a chiral primary \cite{Avery:2010er}. The standard deformation operator takes the form \cite{Guo:2019ady}:
\begin{equation}
\label{eq:Dstandard}
D=\frac14\,\epsilon^{\dot A\dot B}\epsilon_{\alpha\beta}\epsilon_{\dot\alpha\dot\beta}\,G^{\alpha}_{\dot A,-\frac12}\,
\bar G^{\dot\alpha}_{\dot B,-\frac12}\,\twist^{\beta\dot\beta}
=\epsilon^{\dot A\dot B}\,G^{-}_{\dot A,-\frac12}\,\bar G^{-}_{\dot B,-\frac12}\,\twist^{++}\,,
\end{equation}
where in the second step we used $G^{+}_{\dot A,-\frac12}\twist^{-\dot\beta}=-G^{-}_{\dot A,-\frac12}\twist^{+\dot\beta}$, so that $\epsilon_{\alpha\beta}G^\alpha_{\dot A,-\frac12}\twist^{\beta\dot\beta}=-2\,G^{-}_{\dot A,-\frac12}\twist^{+\dot\beta}$, and similarly for the right-movers. The contraction of the indices makes $D$ an R-singlet of weights $(1,1)$, and the contraction of the $SU(2)_2$ indices with $\epsilon^{\dot A\dot B}$ makes it invariant under the rotations of the torus. With $\epsilon^{+-}=-1$ the coefficient matrix $\epsilon^{\dot A\dot B}$ is $- i \sigma_y$, which is the direction we call the singlet. Replacing it by one of the symmetric matrices $1$, $ i \sigma_x$ and $ i \sigma_z$ gives the three components of an $SU(2)_2$ triplet of deformations, which preserves supersymmetry and R-symmetry without being invariant under $SU(2)_2$ \cite{Gaberdiel:2025smz}. The Kummer model has the same structure, with the reflection twist fields at its $16$ fixed points in place of the twist field that exchanges two copies, and we allow the $SU(2)_2$ indices to be contracted with an arbitrary coefficient matrix.\\

\paragraph*{Hermiticity and normalization.}
The general $\mathbb Z_2^4$-symmetric deformation with a complex coefficient matrix $\cMat^{\dot A\dot B}$ is made Hermitian by adding its conjugate:
\begin{equation}
\label{eq:Dherm}
\Dc
 =\frac{1}{2\sqrt{2}} \left[ \cMat^{\dot A\dot B}\,G^{-}_{\dot A,-\frac12}\,\bar G^{-}_{\dot B,-\frac12}\,\twist^{++}  + \cMat^*_{\dot A\dot B}  \;\epsilon^{\dot A\dot C}G^{+}_{\dot C,-\frac12}\;\epsilon^{\dot B\dot D}\bar G^{+}_{\dot D,-\frac12} \;\twist^{--} \right]
 =\frac{1}{2\sqrt{2}} \,(\cMat+\epsilon \cMat^*\epsilon^T)^{\dot A \dot B}\,G^{-}_{\dot A,-\frac12}\,\bar G^{-}_{\dot B,-\frac12}\,\twist^{++}\,,
\end{equation}
where $\cMat^*_{\dot A\dot B} \equiv (\cMat^{\dot A \dot B})^*$ (indices lowered by complex conjugation). This depends only on the matrix $\mathfrak{c} = \frac12(\cMat+\epsilon \cMat^*\epsilon^T)$, which satisfies $\mathfrak{c} ^*=\epsilon\, \mathfrak{c} \,\epsilon^T$. Without loss of generality $\cMat$ is therefore a real quaternion, \eqref{M-eq:quaternion}, and $\Dc$ takes the form \eqref{M-eq:Dc}. An overall phase $\cMat\to e^{ i \theta}\cMat$ would only rescale $\Dc$ by $\cos\theta$, which can be absorbed into $\lam$. The four operators $G^{-}_{\dot A,-\frac12}\bar G^{-}_{\dot B,-\frac12}\twist^{++}$ are orthonormal, since $\twist^{++}$ has unit norm and, by \eqref{eq:GG}, the states $G^{-}_{\dot A,-\frac12}|\twist^{+}\rangle$ have unit norm and are orthogonal for the two values of $\dot A$. Hence the two-point function takes the form:
\begin{equation}
\label{eq:norm}
\langle\Dc(z)\Dc(0)\rangle=\frac{\Tr (\cMat \cMat^\dagger)}{2|z|^{4}}=\frac{|q|^2}{|z|^{4}}\,,
\end{equation}
and we choose $|q|=1$. Every $\Dc$ is exactly marginal, and the deformed theory is the K3 sigma model at a neighboring point of its moduli space, with $c=6$ unchanged \cite{Seiberg:1999xz}.

\section{Symmetries of a direction}
\label{app:symmetries}

\paragraph*{Continuous symmetries.}
The currents $i\partial X$ and $i\bar\partial X$ of the torus, whose charges are the momenta $\ba$ and $\bb$, are odd under the reflection, so the orbifold has already projected them out. In the orbifold the momenta label the untwisted states only up to a common sign and the twisted states do not carry momenta. Among the translations only the half-period shifts $\mathbb Z_2^4$ survive (cf.\ Sec.~\ref{app:kummer}). The weight-one currents of the orbifold are the fermion bilinears, which generate an $su(2)\oplus su(2)$ current algebra at level one in each chirality \cite{Nahm:1999ps,Gaberdiel:2025smz}. On the left-movers, one factor rotates the index $\alpha$ and commutes with the R-singlet $\Dc$. The other factor rotates the index $A$ of the fermions but not that of the bosons, so it does not commute with the supercurrents $G^\alpha_{\dot A}=\psi^{\alpha A}\partial X_{A\dot A}$ and is broken by $\Dc$. Only the two copies of the $\mathcal N=4$ algebra remain as continuous symmetries. These are resolved by working with superconformal primaries of the Ramond sector (Sec.~\ref{app:kummer}).\\

\paragraph*{Mirrors and chambers.}
The equivalences of the full theory described in End Matter act by signed permutations of $(q_0,q_1,q_3)$ and an independent reversal of $q_2$, giving $2^3\,3!\times2=96$ elements. The additional right-moving equivalences of the invariant sector (see below) enlarge this set to all signed permutations of four coordinates, with $2^4\,4!=384$ elements. This is known as the Weyl group $W(B_4)$. Away from the mirrors, the four signs and the ordering of the absolute values specify $384$ ``chambers'' on $S^3$. For example, $q_0\geq q_1\geq q_2\geq q_3\geq0$ defines a fundamental chamber whose four reflecting faces cut out a spherical tetrahedron on $S^3$. Figure~\ref{fig:mirrorgeometry} illustrates the analogous construction on $S^2$, where $2^3\,3!=48$ signed orderings give spherical triangles.\\

\begin{figure}[htb]
\centering
\includegraphics[width=0.8\textwidth]{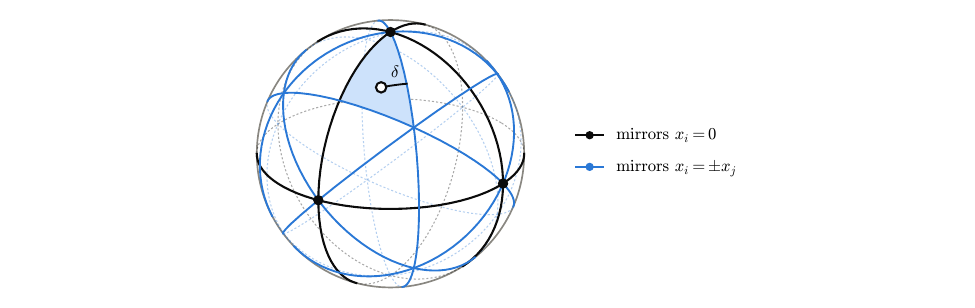}
\caption{Mirrors $x_i=0$ and $x_i=\pm x_{j\neq i}$ on the two-sphere $S^2$. The nine mirrors are great circles, which cut the sphere into $48$ spherical triangles (the chambers of the Weyl group $W(B_3)$). The white dot is a ``generic direction'', at the angle $\delta$ from the nearest mirror.}
\label{fig:mirrorgeometry}
\end{figure}

The singlet deformation \eqref{eq:Dstandard} is the quaternion unit $q=(0,0,-1,0)$, where nine mirrors meet ($q_0=q_1=q_3=0$ and $q_\mu=\pm q_\nu$ for $\mu,\nu\in\{0,1,3\}$). The triplet deformation \cite{Gaberdiel:2025smz} is spanned by the other three quaternion units $(1,0,0,0)$, $(0,1,0,0)$ and $(0,0,0,1)$, i.e., by the mirror $q_2=0$.\\

\paragraph*{The point group and generic directions.}
The square torus $\Tf$ is symmetric under the rotations and reflections of $\mathbb R^4$ that map the lattice $\mathbb Z^4$ to itself. These are $384$ signed permutations of the coordinates, and we consider the orientation-preserving ones. We parametrize the latter as $192$ signed permutation matrices of determinant one. Since the fermions transform as spinors, we then lift the permutations to the double cover $\mathrm{Spin}(4)=SU(2)_1\times SU(2)_2$, obtaining a group $\Gt$ of order $384$. An element $(g_1,g_2)$ acts on quaternion coordinates as $x\mapsto g_1xg_2^{-1}$: $g_1$ rotates the index $A$, and $g_2$ the index $\dot A$. The four central elements $(\pm1,\pm1)$ act trivially on our chosen scalar primaries (by the orbifold projection $(-1,1)$ and even fermion parity $(-1,-1)$).

Only $g_2$ acts on the deformation coefficients. Writing its doublet matrix as $Y$, this action is $\cMat\mapsto Y\cMat Y^T$. It leaves the singlet coordinate $q_2$ unchanged and rotates the triplet $(q_0,q_1,q_3)$ by the $24$ rotations of a cube.

To a given deformation direction $\cMat$ we associate a stabilizer $\Hc$ of point-group operations that preserve the deformation. Operations sending $\cMat \mapsto -\cMat$ also qualify when combined with $\mathcal Q$ (which reverses the sign again). For a generic direction, this requires $g_2=\pm1$ and $g_1\in Q_8 =\{\pm\mathds{1},\pm i\sigma_x,\pm i\sigma_y,\pm i\sigma_z\}$. Thus $\Hc$ has $16$ elements. After quotienting by the four trivial central elements, it acts as $\mathbb Z_2\times\mathbb Z_2$, giving the four real one-dimensional characters used in the main text. Special directions can have larger stabilizers (for example, the singlet preserves all of $\Gt$).

Additional symmetries can involve worldsheet parity, which exchanges the chiralities and sends $\cMat\mapsto\cMat^T$, reversing $q_2$. It is therefore a symmetry on the mirror $q_2=0$. Its compositions with point-group operations and with ${\cal Q}$ give the symmetries on the other nine mirrors that exist in the full theory.

The six additional mirrors $q_2=\pm q_\mu$ arise from rotations of the right-moving fields alone by $(u,u)$, with $u\in\{i\sigma_x,i\sigma_y,i\sigma_z\}$. These send $\bb\mapsto u\bb u^{-1}$, changing coordinate signs and therefore preserving $\ba\equiv\bb\pmod2$ in the invariant sector. They commute with $H_0$ and preserve both $\mathcal N=4$ algebras, rotating the right-moving supercurrents among themselves. Covariance of the right-moving chiral twist maps gives $\Dcz\mapsto\mathcal D_{\cMat u,0}$ on the entire invariant sector, including matrix elements between different levels. The spectra along $\cMat$ and $\cMat u$ therefore coincide beyond first order.

Composing these rotations with the equivalences of the full theory produces the reflections in the six additional mirrors. These rotations do not preserve the full momentum-winding lattice, so the additional equivalences do not extend to the full theory. Preserving the integer lattice and its parity pairing for every allowed momentum pair restricts these right-moving rotations to $u\in Q_8$.

One further ``hidden'' $\mathbb Z_2$ of the even levels, the same along every direction, is described in End Matter; we give its explicit form at $E=2$ below.\\

\paragraph*{The hidden $\mathbb Z_2$ at $E=2$.}
The hidden $\mathbb Z_2$ (End Matter) does not depend on the direction: an involution $S$ can be found, which commutes with $H_0$, with $\Dcz$ for every $\cMat$, and with every element of $\Gt$. Along the singlet it therefore acts within each sector of $\Gt$, and it can be written down explicitly at the lowest level. At $E=2$, the sector that transforms trivially under $\Gt$ and has $Z_R=+1$ contains one twisted state and three untwisted states: the invariant oscillator state $|{\rm osc}\rangle$ ($\ba=\bb=0$); the combination $|{\rm m}{+}{\rm w}\rangle=\frac1{\sqrt2}\big(|{\rm m}\rangle+|{\rm w}\rangle\big)$ of the invariant pure-momentum state $|{\rm m}\rangle$ ($\bb=\ba$, $|\ba|^2=2$) and pure-winding state $|{\rm w}\rangle$ ($\bb=-\ba$), which are exchanged by $Z_R$; and the invariant state $|{\rm mw}\rangle$ with $|\ba|^2=|\bb|^2=2$, $\bb\neq\pm\ba$. For suitable phases of these normalized states, $S$ acts on them as the reflection
\begin{equation}\label{eq:Sreflection}
S=\frac{2\,|v\rangle\langle v|}{\langle v|v\rangle}-1\,,\qquad |v\rangle=2\,|{\rm osc}\rangle+|{\rm m}{+}{\rm w}\rangle-|{\rm mw}\rangle\,,
\end{equation}
and as $-1$ on the twisted state (the overall sign of $S$ is a convention). Along the singlet, $|v\rangle$ therefore stays at $E=2$ to first order. Note that $S$ mixes an oscillator state with momentum states of the same energy, which no automorphism of the momentum lattice can do; it is therefore not in the point group, and it is not $Z_R$, worldsheet parity or $\mathcal Q$.\\


\paragraph*{The antiunitary symmetry and symmetry class.}
Since $\Dc$ is a Hermitian scalar, CPT maps $\Dc(\sigma,0)$ to $\Dc(-\sigma,0)$ and therefore commutes with its spatial integral $\Dcz$. Composing CPT with the Weyl reflections of both R-symmetry groups returns the R-charges to our chosen lowest weights and defines an antiunitary operator $T$. The Weyl reflections preserve the R-singlet deformation, so
\begin{equation}
T H_0 T^{-1}=H_0\,,\qquad
T\Dcz T^{-1}=\Dcz \,.
\end{equation}
In the free field realization, $T$ conjugates the coefficients of Fock monomials, acts by $i\sigma_y$ on each internal index $A$ and $\dot A$, and it reverses the momenta, leaving the R-symmetry indices unchanged. Since $(i\sigma_y)^2=-1$, its square is $+1$ on bosonic oscillators, which carry both internal indices, and $-1$ on fermionic oscillators, which carry only $A$. The chosen untwisted Ramond vacuum and the twisted ground states are preserved up to phases, so $T^2=+1$ on these states. Therefore,
\begin{equation}
T^2=(-1)^F=+1
\end{equation}
on our selected space: the fixed left- and right-moving R-charges require even total fermion parity in both sectors.

The operator $T$ commutes with $\Hc$, $Z_R$ and $\mathcal Q$. For a generic direction, the characters of $\Hc$ are real, so $T$ preserves each character space and each $Z_R$ eigenspace. Where the hidden $\mathbb Z_2$ is resolved numerically (End Matter), the two jointly invariant subspaces have unequal dimensions. An antiunitary symmetry commuting with both $H_0$ and $\Dcz$ can only permute their irreducible components. Since the two components have unequal dimensions, $T$ must preserve each separately.

Since $T^2=+1$, we can choose an orthonormal basis in which $T$ acts by complex conjugation. Each fully resolved first-order matrix along a generic direction is then real symmetric. Moreover, $[T,\mathcal Q]=0$ and $\{\mathcal Q,\HB^{(1)}\}=0$ imply that the antiunitary operator $C=\mathcal Q T$ satisfies
\begin{equation}
\label{eq:Csquared}
C^2=+1\,,\qquad C\HB^{(1)}C^{-1}=-\HB^{(1)}\,.
\end{equation}
Together, these symmetries place the first-order splitting matrices $\HB^{(1)}$ in class BDI, the chiral GOE \cite{Verbaarschot:1994qf,Altland:1997zz}. Agreement of their non-zero bulk spectra with GOE statistics is tested numerically in Sec.~\ref{M-sec:firstorder}.

\section{First-order matrix and spectral statistics}
\label{app:firstorder}

\paragraph*{Relation to the symmetric orbifold.}
The twist-field construction is inherited from $\Sym^2(\Tf)$, the smallest non-trivial member of the D1-D5 symmetric orbifold family \cite{Larsen:1999uk,David:2002wn,Lunin:2001pw,Keller:2011xi,Hartman:2014oaa}. For concreteness, introduce the canonically normalized fields
\begin{equation}
X_{\rm cm}=\frac{1}{\sqrt{2}} \left(X^{(1)}+X^{(2)}\right)\,, \qquad
X_{\rm rel}=\frac{1}{\sqrt{2}} \left(X^{(1)}-X^{(2)}\right)\,.
\end{equation}
Exchange of copies preserves $X_{\rm cm}$ and reflects $X_{\rm rel}$, although their compactification lattice correlates the two. In the untwisted sector with vanishing center-of-mass momentum and winding, write $\bn_1=-\bn_2\equiv\bn$ and $\bw_1=-\bw_2\equiv\bw$.
For copies at $R_{\rm copy}=\sqrt2$ (in units $\alpha'=2$) the relative momenta are
\begin{equation}
\ba=\sqrt2\,\mathbf p_L^{(1)}=\bn+\bw\,, \qquad
\bb=\sqrt2\,\mathbf p_R^{(1)}=\bn-\bw \,.
\end{equation}
This is the invariant lattice of the radius-$2$ Kummer model, with $\bn_{\rm K}=2\bn$ and $\bw_{\rm K}=\bw$. For each choice of a center-of-mass Ramond ground state, the relative states form one copy of the invariant Kummer sector ${\cal H}_{\bs=0}$, including its twisted tower. Related twist deformation calculations through second order include
\cite{Gava:2002xb,Avery:2010er,Gaberdiel:2015uca,Keller:2019yrr,Keller:2019suk,Guo:2019ady,Guo:2020gxm}.\\

\paragraph*{Matrix elements.}
Since $\Dcz$ conserves the spin, the $\sigma$-integral in its definition gives:
\begin{equation}
\label{eq:psiprimeDpsi}
\langle\Psi_b|\Dcz|\Psi_a\rangle= \left\{ \begin{aligned} & 2\pi\,\langle\Psi_b|\Dc(0,0)|\Psi_a\rangle && \text{if } h_{\Psi_a}-\bar{h}_{\Psi_a} = h_{\Psi_b}-\bar{h}_{\Psi_b}\,, \\
&0 && \text{otherwise}\,.
\end{aligned} \right.
\end{equation}
Between two spin-zero states of the same level all four weights equal $\frac E2$. Off the level diagonal $\Dcz$ changes $E$, which matters only beyond first order. The chiral twist map ${\cal A}_{\dot A}$ sends an untwisted chiral state to the twisted chiral state produced by $G^{-}_{\dot A,-\frac12}\twist^{+}$, with $\twist^{+}$ the left-moving factor of $\twist^{++}$. A $\mathbb Z_2$ twist field reflects the fields along a loop around its insertion. On the double cover of the worldsheet branched at that point the reflected field is single-valued \cite{Dixon:1986qv}, so that a twisted-sector matrix element becomes a free-field computation on the cover. In the form of this construction developed for the symmetric orbifold \cite{Lunin:2001pw,Avery:2010er,Guo:2019ady}, the two sheets of the cover are the two copies of $\Tf$ in $\Sym^2(\Tf)$, and the reflection twist field of the Kummer torus is their exchange.

The level-preserving part of ${\cal A}_{\dot A}$ commutes with every mode of the chiral $\mathcal N=4$ algebra, zero modes included, and therefore maps massive superconformal primaries into primaries and each descendant into the corresponding descendant. The level-preserving mode of ${\cal X}=G^-_{\dot A,-\frac12}\twist^+$ is ${\cal X}_0$ (labeled by how much it lowers $L_0$). The operator product of ${\cal X}$ with any $\mathcal N=4$ generator $W$ is a total derivative, $W(w){\cal X}(z)\sim\partial_z[\Phi_W(z)/(w-z)]$, so that $[W_n,{\cal X}_k]=-k\,\Phi_{W,n+k}$, which vanishes for $k=0$. Here $\Phi_T={\cal X}$ since ${\cal X}$ is a primary of weight one, $\Phi_{J^a}=0$ since ${\cal X}$ is an R-singlet primary, and $\Phi_{G^\pm_{\dot B}}\propto\epsilon_{\dot B\dot A}\twist^\pm$ by \eqref{eq:GG}, since $\twist^+$ is a chiral primary and ${\cal X}$ an R-singlet.

To measure the statistics at a level $E$, we build $\Mc$ on the complete space of massive primaries at that level and project it onto the multiplicity spaces of the physical irreducible representations of $\Hc$. We split each of them by $Z_R$ and, at the even levels, by the hidden $\mathbb Z_2$ (End Matter). Along the singlet, $\cMat^T=-\cMat$, so worldsheet parity combined with $\mathcal Q$ is a further symmetry (Sec.~\ref{app:symmetries}), which also leads to a further splitting. In each fully resolved sub-block we retain the distinct non-zero singular values. Working in multiplicity spaces removes degeneracies due to the symmetry representations, and we additionally count any remaining exactly degenerate value only once. In our data, such additional degeneracies occur only along the singlet (Sec.~\ref{M-sec:firstorder}). We use the same distinct values for the gap ratios and for the spacing statistics. Keeping the singlet's degenerate values with their multiplicity instead makes $9\%$ of its unfolded spacings at $E=8$ exactly zero and lets the few large multiplets dominate $\Sigma^2(L)$.\\

\paragraph*{Unfolding.}
In a fully resolved sub-block, let $\nu_i$ be the distinct ordered singular values and $N(\nu)$ the number of them not larger than $\nu$ (the counting function). The smooth part $\bar N(\nu)$ of this staircase captures the mean density, which is not universal and changes from one sub-block to another, while RMT predicts the fluctuations around it. Unfolding replaces the singular values by the unfolded levels:
\begin{equation}
\label{eq:unfold}
x_i=\bar N(\nu_i)\,,\qquad i=1,\dots,n\,,
\end{equation}
which have unit mean density everywhere. In practice, we take $\bar N$ to be the least-squares fit of $N(\nu_i)=i$ by a Chebyshev polynomial in $\nu$, and we drop $5\%$ of the values at each end of the band, where the fit is least reliable. The degree of the polynomial grows with the size of the sub-block: for $n$ levels it is $n/100$ rounded to the nearest integer, and at least $12$. With one fit parameter per $100$ levels, $\bar N$ follows the mean density on scales much longer than the intervals $L\le10$ considered below. A polynomial of fixed low degree is not enough at the higher levels: at degree $12$ the fit of the larger sub-blocks leaves a residual that varies slowly across the band, which raises the number variance at large $L$. As a control, we unfolded chiral GOE spectra with the shapes of the reference direction's sub-blocks in the same way. They reproduce the GOE number variance for $L\le10$ at $E=6,7,8$, so the fit does not absorb the fluctuations on these scales. At $E=6$ the sub-blocks have between $451$ and $1366$ levels, and the fit cannot be refined further without absorbing them. Sub-blocks are unfolded separately, since their levels are uncorrelated. A sub-block is used only if the fitted staircase is monotone over the retained range, since otherwise unfolded spacings can become negative. Along the singlet direction, whose level density has sharp peaks, too few sub-blocks achieve this for $E\leq 6$, so Fig.~\ref{M-fig:spacings} shows its unfolded statistics only for $E\geq7$.\\

\paragraph*{Spacing distribution and number variance.}
The unfolded nearest-neighbor spacings $s_i=x_{i+1}-x_i$ have the distribution $P(s)$. Unlike the gap ratio, it depends on the unfolding. The Wigner surmises for the Gaussian ensembles and the distribution for uncorrelated levels read \cite{Mehta2004}:
\begin{equation}
\label{eq:wigner}
P_{\rm GOE}(s)=\frac{\pi s}{2}\,e^{-\pi s^2/4}\,,\qquad
P_{\rm GUE}(s)=\frac{32\,s^2}{\pi^2}\,e^{-4s^2/\pi}\,,\qquad
P_{\rm Poisson}(s)=e^{-s}\,.
\end{equation}
Let $n(x,L)$ be the number of unfolded levels in the interval $[x,x+L]$. Its mean is $L$ and its variance over the position of the interval,
\begin{equation}
\label{eq:sigma2}
\Sigma^2(L)=\left\langle n(x,L)^2\right\rangle_x-\left\langle n(x,L)\right\rangle_x^2\,,
\end{equation}
measures how rigid the spectrum is over $L$ levels. The universal predictions are \cite{Mehta2004}:
\begin{equation}
\label{eq:sigma2goe}
\begin{aligned}
\Sigma^2_{\rm GUE}(L)&=\frac{1}{\pi^2}\Big[\log(2\pi L)+\gamma+1-\cos(2\pi L)-\mathrm{Ci}(2\pi L)\Big]
+L\Big(1-\frac{2}{\pi}\,\mathrm{Si}(2\pi L)\Big) \, \stackrel{L\gg 1}{\approx}\, \frac{1}{\pi^2} \,\log(L) + 0.346 \,,\\
\Sigma^2_{\rm GOE}(L)&=2\,\Sigma^2_{\rm GUE}(L)+\frac{\mathrm{Si}(\pi L)^2}{\pi^2}-\frac{\mathrm{Si}(\pi L)}{\pi}  \, \stackrel{L\gg 1}{\approx} \, \frac{2}{\pi^2}  \, \log(L) + 0.442
\,, \qquad\qquad
\Sigma^2_{\rm Poisson}(L) =L \,,
\end{aligned}
\end{equation}
where $\mathrm{Si}$ and $\mathrm{Ci}$ are the sine and cosine integrals and $\gamma$ is the Euler--Mascheroni constant. While $P(s)$ and the gap ratio probe neighboring levels, $\Sigma^2(L)$ probes correlations on larger scales. All three statistics are computed within each sub-block, and then combined: the pooled $\rmean$ and $P(s)$ count every ratio or spacing once, so that each sub-block enters in proportion to its size, and $\Sigma^2(L)$ is the average of the sub-blocks' number variances weighted by their numbers of levels. In Fig.~\ref{M-fig:spacings} (left), the reference direction contributes $33157$ unfolded spacings at $E=8$ from its five fully resolved sub-blocks, and the singlet $5704$ spacings from its fully resolved sub-blocks, with each degenerate value counted once.\\


\ifdefined\arXivversion\else\input{\mainDocFilename.bbl}\fi

\fi

\end{document}